\documentclass[aps,prb,groupedaddress,onecolumn]{revtex4}
\usepackage{graphicx}

\begin{document}

\title{Single donor ionization energies in a nanoscale CMOS channel}
\author{M. Pierre}
\author{R. Wacquez}
\author{X. Jehl}
\author{M. Sanquer}
\affiliation{INAC-SPSMS, CEA-Grenoble, France.}
\author{M. Vinet}
\author{O. Cueto}
\affiliation{CEA/LETI-MINATEC, CEA-Grenoble, France.}

\maketitle

{\bf One consequence of the continued downwards scaling of transistors is the reliance on only a few discrete atoms to dope the channel, and random fluctuations of the number of these dopants is already a major issue in the microelectonics industry \cite{Intel45nm}. While single-dopant signatures have been observed at low temperature  \cite{Schofield03,Simmons06,Zhong05,Ono07,Khalafalla07,Khalafalla09,Delft08}, studying the impact of only one dopant up to room temperature requires extremely small lengths. Here, we show that a single arsenic dopant dramatically affects the off-state behavior of an advanced microelectronics field effect transistor (FET) at room temperature. Furthermore, the ionization energy of this dopant should be profoundly modified by the close proximity of materials with a different dielectric constant than the host semiconductor \cite{Diarra07,Bjork09}. We measure a strong enhancement, from 54\,meV to 108\,meV, of the ionization energy of an arsenic atom located near the buried oxide. This enhancement is responsible for the large current below threshold at room temperature and therefore explains the large variability in these ultra-scaled transistors. The results also suggest a path to incorporating quantum functionalities into silicon CMOS devices through manipulation of single donor orbitals.}



Progresses in nanotechnology made it possible to observe signatures of single dopants in semiconducting nanostructures, with for instance STM technique\cite{Schofield03,Simmons06} or grown nanowires \cite{Zhong05}. Single dopants have been detected as a hump in the sub-threshold slope of a capacitively coupled surface channel \cite{Ono07,Khalafalla07,Khalafalla09}. Resonant tunneling transport via an impurity state has been studied in structures markedly larger than a typical donor orbital \cite{Bending85,Fowler86,Geim94,Savchenko95,Calvet08}. Resonant transport spectroscopy has even been performed in a transistor geometry favoring a strong hybridization of a donor electronic wavefunction with a surface channel \cite{Delft08}.
Both a very small volume and a small number of dopants in the channel are required to observe resonant transport through the first isolated dopant. Otherwise hybridization with surface states, polarization of nearby dopants, or many-dopant problem \cite{kuznetsov97} will dramatically modify the wavefunction and energy spectrum. The distance between the source and drain is also of crucial importance because the barriers will have an exponential dependence on this length divided by the characteristic wavefunction extension. 

\begin{figure*}[!t]
\begin{center}
\includegraphics[width=0.85\textwidth]{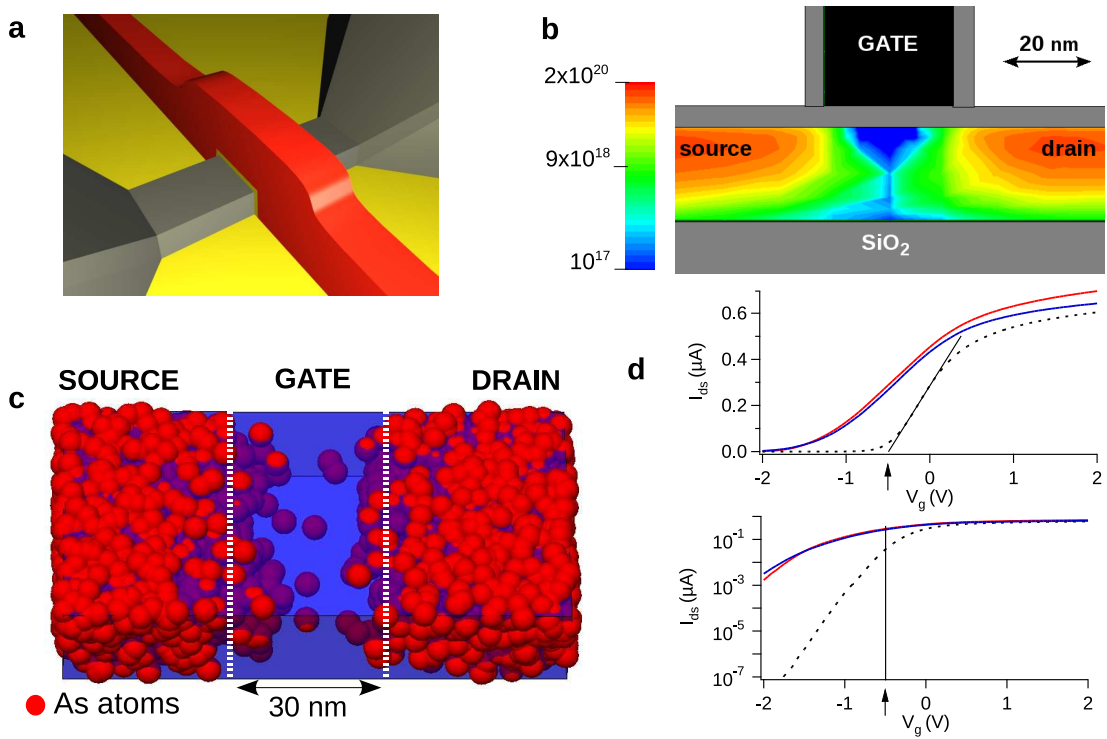}
\caption{{\bf Geometry, simulations and electrical characteristics of the devices. a,} Scaled schematic view of a 30\,nm gate length sample. {\bf b,} Simulation of the doping concentration (color scale in cm$^{-3}$) along the channel, in a continuous approximation, showing that a maximum if reached near the center of the film. {\bf c,} Simulation with discrete As dopants, represented by red spheres of radius the mean bare Bohr radius (2.2\,nm). The very few isolated dopants in the center are responsible for the large off-state current observed in some samples. The dotted lines are the edges of the gate, separated by 30\,nm. {\bf d,} Electrical I$_{\mathrm d}$-V$_{\mathrm g}$ characteristics of 3 nominally identical samples at 300\,K, measured with V$_{\mathrm d}$=10\,mV, showing in solid lines two devices with large off-state current at small V$_{\mathrm g}$ and one with small off-state current (dashed line). The threshold voltage is -0.5\,V, indicated by the arrows.}
\label{fig1}
\end{center}
\end{figure*}

Therefore a very natural way to study transport through only one dopant is to take advantage of CMOS technology to build a very short, narrow and doped transistor. We fabricated silicon-on-insulator (SOI) transistors (Fig. \ref{fig1}a) with a geometry and process designed to take advantage of the inhomogeneous dopant profile after ion-implantation of the source and drain. A simulation of this doping profile, based on the real process and continuous approximation, is shown in Fig. \ref{fig1}b. A similar simulation taking into account discrete dopants with a Kinetic Monte Carlo approach for diffusion is shown in Fig. \ref{fig1}c, with the As atom size set to the mean Bohr radius in bulk Si, $a_{B}=2.2\,\mathrm{nm}$. All the regions with a concentration above the Mott transition of bulk Si, that is $n_{c} = 8\times 10^{18} As\,\mathrm{cm^{-3}}$ [\onlinecite{holcomb}], must be considered as part of the source and drain. Based on these simulations the effective electrical channel length is of the order of 10\,nm. Therefore the remaining volume available for As dopants below $n_c$ is roughly $V=$10$\times$20$\times$50\,nm$\mathrm{^3}$. The probabilities of finding one dopant near the buried oxide (for maximum ionization energy enhancement) or a dopant cluster, most likely a donor pair, can be evaluated and compared. For $N$ dopants the probability of finding a pair with separation $a_{B}$ or less is given by \cite{Geim94}: $$ P_{pair}=1-\prod_{n=1}^N{(1-nz)}$$ where $z$ is the ratio $\frac{4}{3}\pi {a_{B}}^3/V\approx0.005$ . For 5 dopants $P_{pair}$ is only 5\%, but it reaches 90\% for 30 dopants. In the same time the probability for having at least one donor out of five at 1\,nm or less from the BOX interface is already 25\%.

These ultra-scaled dimensions, corresponding to a 10\,nm technology node transistor \cite{Intel45nm}, naturally yield variability. A set of 25 nominally identical samples was measured. We found similar on-state current at saturation (V$_{\mathrm g}$=+2\,V) but very large sub-threshold dispersion (see Supplementary Information, Fig. S1). Two extreme room temperature characteristics, featuring the highest and lowest off-state currents (at V$_{\mathrm g}$=-2\,V) are shown in Fig.\ref{fig1}d. A threshold voltage of -0.5\,V $\pm 0.05$ is obtained after extrapolating the point of maximum transconductance in the linear I$_{\mathrm d}$-V$_{\mathrm g}$ characteristics of the samples with the lowest off-state current and steepest subthreshold swing (arrows in Fig. 1d). This phenomenological criterion commonly used in microelectronics is well adapted to such mesoscopic samples where the usual definitions (relying on self-averaging) are not valid. At T=4.2\,K these samples exhibit a pattern of large drain current oscillations above V$_{\mathrm g}$=-0.2\,V and the drain conductance quickly increases up to or above the quantum of conductance $\frac{e^2}{h}\approx 25.8\,k\Omega$ that fixes the onset of a diffusive channel at low temperature (not shown). The shift of the onset voltage from $-0.5\,V$ to $-0.2\,V$ is due to localization of carriers at low temperature. Above V$_{\mathrm g} \simeq$-0.2\,V resonant states observed at low temperature result from hybridization of several electronic states, for instance a donor state hybridized with a surface state\cite{Delft08}.

\begin{figure*}
\begin{center}
\includegraphics[width=0.85\textwidth]{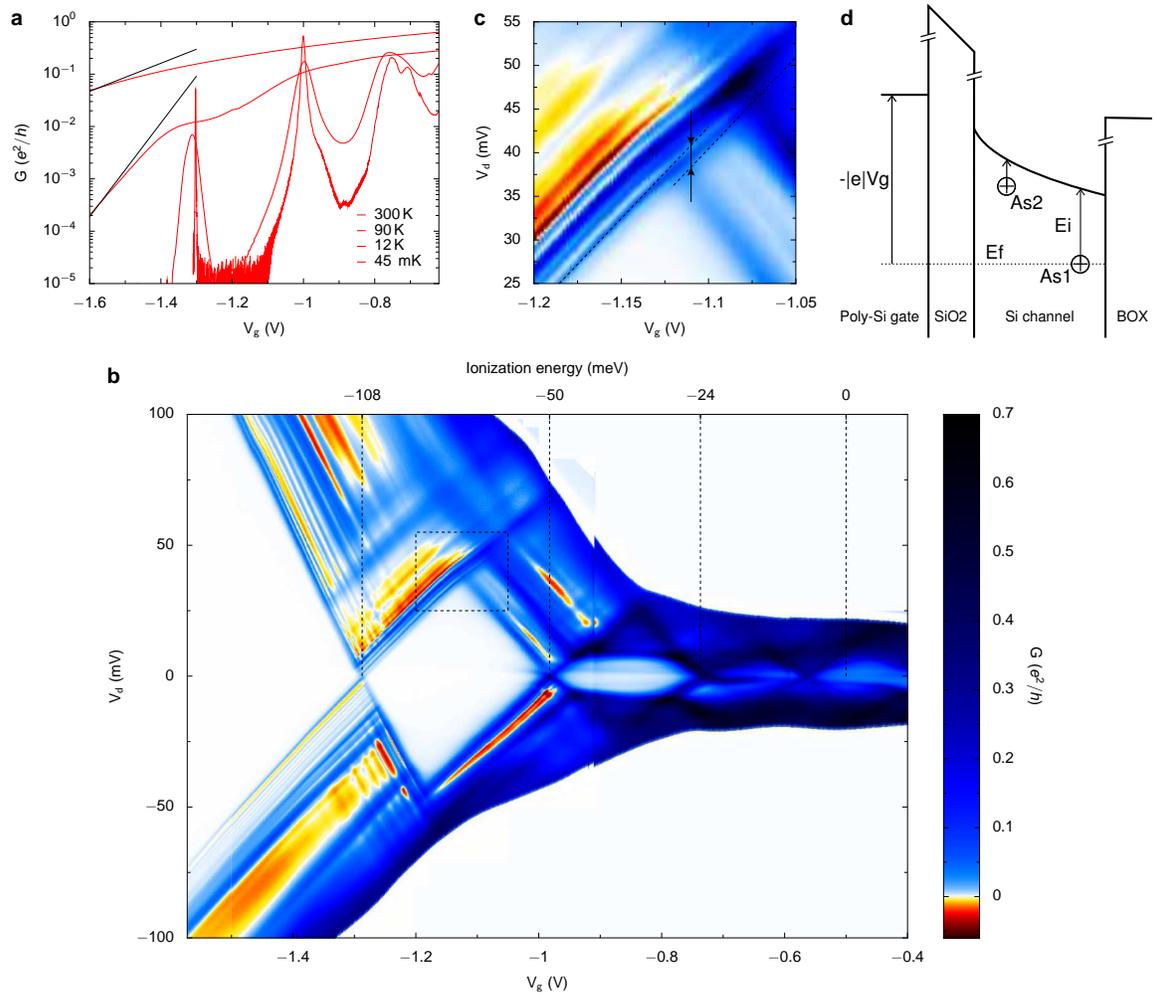}
\caption{{\bf Differential source-drain conductance versus gate and drain voltages at 4.2\,K for the sample with the largest off-state current (blue line in figure \ref{fig1}d). a,} Differentiel source-drain conductance versus gate voltage at various temperatures. The black lines account for thermal activation to the first resonance. {\bf b}, The top horizontal scale is the energy in the channel counted from the threshold voltage. The first peak corresponds to the first As dopant with a strongly enhanced ionization energy due to dielectric confinement near the buried oxide interface. The second and third resonances correspond to different electronic occupations of another donor. All the negative differential conductance lines (in orange) are due to local density-of-states fluctuations in the source/drain (see figure \ref{1erpic}). Only two positive differential conductance lines (one for each polarity) are attributed to an excited state of the second peak (Fig. \ref{cotunneling}). {\bf c}, Detail of the dotted box area shown in panel {\bf b} where the first and second diamonds cross each other without being strongly affected. The slight shift observed for the first diamond and highlighted by the arrow corresponds to the shift of the first donor's energy due to the modification of the static potential created by the second donor being filled ($ \simeq $ -2.5\,meV). {\bf d}, Simplified energy diagram across the structure, from the gate down to the buried oxide (BOX). Two As donors are represented, the one closer to the BOX has a larger ionization energy and is probed at large negative values in $V_g$. }
\label{rhombus4K}
\end{center}
\end{figure*}

\begin{figure}
\begin{center}
\includegraphics[width=0.6\columnwidth]{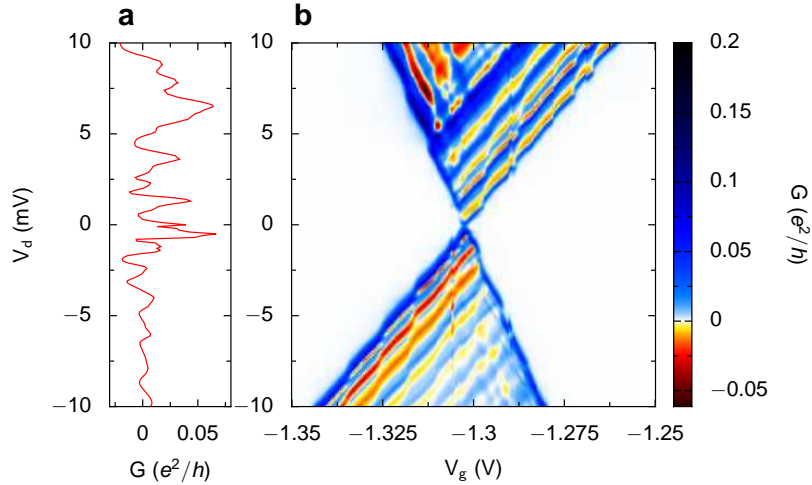}
\caption{{\bf Coulomb blockade spectroscopy of the first dopant atom. a,} Differential conductance versus bias voltage V$\mathrm{_d}$ at $V_{\mathrm g}$=-1.3\,V. {\bf b,} Close-up of the first peak of Figure \ref{rhombus4K}, recorded at T=100\,mK. The differential conductance $ \propto {d \nu_{D}\over dE}$ (parallel to the right edge) or $ \propto {d \nu_{s}\over dE}$ (parallel to the left edge) depending on the exact balance between tunneling rates to source and drain. The typical 1\,meV correlation energy for the lines is the inverse diffusion time in the electrodes before an inelastic event (corresponding diffusion length about 10\,nm).}
\label{1erpic}
\end{center}
\end{figure}

New and strikingly different features are observed at low temperature for the samples with large off-state current at 300\,K. A very few resonances appear at gate voltages down to -1.3\,V, i.e.\ when electrons are pushed away far from the gate. A 2D plot of differential conductance at 4.2\,K versus drain-source voltage is shown in Fig. \ref{rhombus4K}. The gate voltage is translated in energy (referenced from the band edge) using the lever arm parameter $\alpha$=0.16 extracted from the Coulomb diamonds. This rather small value for $\alpha$ indicates that the orbital responsible for this resonance is less coupled to the gate than to source and drain, i.e.\, in our geometry, that the electronic state is rather on the BOX side of the nanowire.
The value $V_{g1}=-1.3\,V$ combined with $\alpha$=0.16 and a threshold voltage of -0.5\,V yields an ionization energy for the first donor of 108\,meV$\pm$10\,meV, a value markedly larger than the 53.7\,meV expected \cite{Kohn55} for bulk Si. Note that the other sample shown in Fig. \ref{fig1}d shows a value 98\,meV$\pm$10\,meV, i.e.\ also a strong enhancement.
Fig. \ref{rhombus4K}a shows that the drain current in excess at low V$_{\mathrm g}$ and temperature up to 300\,K  is due to thermal broadening of these resonances. For instance for $V_{g}\le V_{g1}$ the current is given by thermal activation to the first resonance: $ I_{ds}\propto exp(-(e\alpha (V_{g}-V_{g1})/k_{B}T$). The sub-threshold variability is the most affected by the presence of a single centered donor, which increases the drain conductance by four orders of magnitude at V$_{\mathrm g} $=-1.5\,V and at room temperature (see Fig.\ref{fig1}d), a much larger effect than reported before \cite{Ono07}.
Detailed transport spectroscopy of the first resonance  is shown in Fig. \ref{1erpic}. No electronic states exist below $V_{g1}$ since no distortion of the diamond occurs at large drain voltage, up to 100\,mV (see Fig. \ref{rhombus4K}). The absence of electrons in the channel at lower energy is crucial to ensure that the spectroscopy of the state at $V_{g1}$ is not modified by electronic correlations with other electrons weakly bound on extra donors (D$^{0}$ states). We observe many lines of differential conductance parallel to both edges of the diamond and successively positive and negative (see Fig. \ref{1erpic}), due to fluctuations of the local density-of-states (LDOS) in the source and drain \cite{Falko97,Falko01}. Lines with different slopes correspond to probing the source or the drain, depending on the exact balance of the tunneling rates (see Supplementary Information, Fig. S2). The large observed LDOS fluctuations originate from the extremely small non-invasive As doped contacts with a finite transverse doping gradient. There is no evidence of any differential conductance line due to an excited state of the donor at least up to $V_d$=40\,mV. 



Donor states in bulk silicon are well described \cite{Kohn55} by a hydrogenic model extended by variational methods to account for the anisotropy of the effective mass of Si. The ground state lies 53.7\,meV below the conduction band and the first excited state at 32.6\,meV. More recently strong corrections have been predicted for silicon nanowires, either because of quantum confinement, relevant for diameters below 5\,nm [\onlinecite{Niquet06}], or due to dielectric confinement, more important for larger diameters \cite{Diarra07}. Compared to bulk Si case, an enhanced ionization energy is expected for donors close to a Si/SiO$\mathrm{_2}$ interface, and a reduction for dopants close to the gate, due to the long range screening of the positive core donor charge potential for electrons. This has a dramatic impact on dopant ionization even at room temperature \cite{Diarra07,Bjork09}.
In our data the first resonance occuring 108\,meV below the conduction band (Fig. \ref{rhombus4K}) illustrates the sensitivity of the ionization energy with the dielectric environment. Following ref. \onlinecite{Diarra07}, we can estimate the correction to the ionization energy for a single dopant at position $r_{0}$ and distance $z$ from the dielectric interface as: $$\Delta E_{I} \simeq \langle \psi \mid V_{s}(r,r_{0}) \mid \psi \rangle,$$ where $V_{s}(r,r_{o})$ is the image potential at position $r$ and $\psi$ the bound state on the donor. The approximation $V_{s}(r,r_{o}) \approx V_{s}(r_{o},r_{o})=V_{s}(z)$ yields 
$$\Delta E_{I} \approx  \frac{1}{4 \pi \epsilon_{0}} \frac{e^{2}}{2z} \frac{\epsilon_{\mathrm{Si}}-\epsilon_{\mathrm{SiO_2}}}{\epsilon_{\mathrm{Si}}(\epsilon_{\mathrm{Si}}+\epsilon_{\mathrm{SiO_2}})} \simeq \frac{30.5\,\mathrm{meV}}{z\,(\mathrm{nm})}$$

\begin{figure}
\begin{center}
\includegraphics[width=0.6\columnwidth]{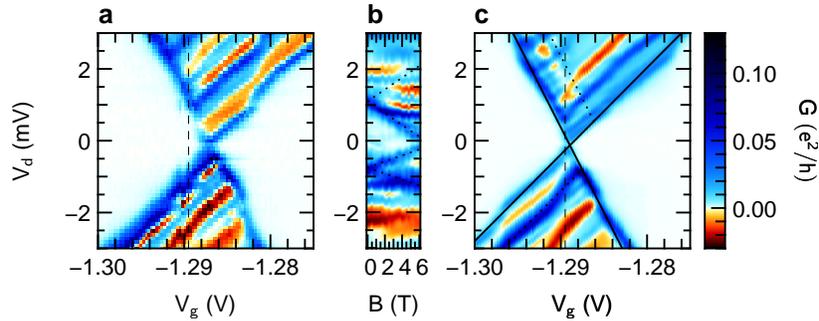}
\caption{{\bf Magnetic field dependence of the first peak. a,} Coulomb diamond spectroscopy at zero field. {\bf b,} Magnetic field dependence at V$\mathrm{_g}$=-1.2895\,V (dashed line in a), the magnetic field being applied parallel to the wire. {\bf c,} Coulomb diamond at 6\,T. The dotted lines in {\bf b} are the predicted Zeeman shifts for $\Delta$S$_z=\pm \frac{1}{2}$ and g=2. The different slopes correspond to different level arm factors extracted from the Coulomb diamonds (no adjustable parameter). For negative drain voltage the positive differential conductance line associated to the spin up level is clearly seen. For positive drain voltage the corresponding line crosses negative differential conductance lines due to the spectroscopy of the drain, resulting in a cancellation of the differential conductance rather than a positive conductance line.}
\label{champ}
\end{center}
\end{figure}

The effects of screening by the source and drain, as well as quantum confinement \cite{Niquet06} are neglected. 
This estimation indicates that a large effect as we observe is realistic for a dopant very near the buried oxide interface ($\leq$1\,nm). Theory also predicts an extension of the bound state function along the nanowire axis and a reduction in the transverse direction \cite{Diarra07}. This should increase the energy difference between the ground and first excited state. Indeed our experiment gives a lower bound of 40\,meV for this energy, significantly larger than the expected 21\,meV for As in bulk Si. The shift of electrostatic potential $V_{s}(z)$ due to the positive image charge in the BOX is nearly equivalent to the bare Coulomb potential of another ionized donor located at $4z$ in bulk silicon. The mean distance between As donors is 20\,nm resulting in a bare shift of $\Delta E_{I}(20\,\mathrm{nm}) \simeq 6\,\mathrm{meV}$ for each extra ionized donor. However donors are located less than $d$=5\,nm from either the source or drain which efficiently screen the bare potential by a factor $2(\frac{d}{z})^2$ [\onlinecite{sanquer00}]. $\Delta E_{I}(20\,\mathrm{nm})$ is then reduced to less than 1\,meV, a value much smaller than the observed energy enhancement. The shift due to the ionization of the second donor is in fact directly measured ($\simeq$-2.5\,meV) as the shift of the first diamond as it crosses the second one, which corresponds to the $D^{+} \rightarrow D^{0}$ transition for the second donor (see Fig. \ref{rhombus4K}c).
Fig. \ref{champ} shows the evolution of the first peak with parallel magnetic field. A Zeeman splitting with a Land\'e factor g=2 is observed, both as a shift of the ground state down to lower energy and as an extra line of differential conductance (no adjustable parameter). This is in agreement with our interpretation of the first peak corresponding to the $ D^{+} \rightarrow  D^{0} $ transition, i.e.\ a spin change of $\frac{1}{2}$.

\begin{figure}
\begin{center}
\includegraphics[width=0.6\columnwidth]{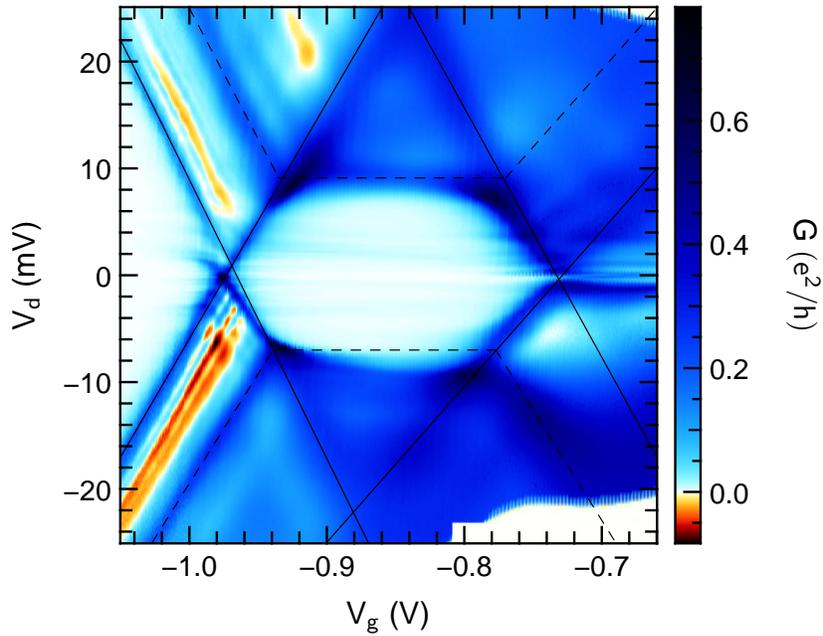}
\caption{{\bf The second and third peaks of Fig. \ref{rhombus4K} recorded at T=100\,mK,} respectively corresponding to the $D^{+} \rightarrow D^{0}$ transition (second) and $D^{0} \rightarrow D^{-}$ (third) . Solid lines indicate the diamond edges while dashed lines highlight features due to inelastic co-tunneling via the first excited state of the  donor. The measured difference in energy between the ground and first excited state is 8\,meV, and the charging energy is 18\,meV. Lines inside the diamonds are due to LDOS in the electrodes revealed by elastic cotunneling.}
\label{cotunneling}
\end{center}
\end{figure}

The second resonance appears near V$\mathrm{_g}$=-1\,V, i.e.\ $\approx$\,50\,meV below the conduction band. Because the lever arm parameter is different from the first resonance and the lines of differential conductance cross the first diamond with only a slight modification (see Fig. \ref{rhombus4K}c), the second resonance does not correspond to the double occupation of the first As donor. Indeed confinement makes it very unlikely for a second electron to sit on the same dopant. The corresponding charging energy is certainly larger than the ionization energy of another As atom unavoidably present in the channel. Therefore the second peak corresponds to a second dopant located closer to the gate oxide as it is sensed at higher gate voltage. This should have four main consequences: $\Delta E_{I}$ should be strongly reduced, the excited state should be lower in energy, double occupancy could be observed (albeit very close to the conduction band \cite{Delft08}), and the tunneling rate is likely to be strongly enhanced due to a larger orbital extension.
All these features are observed, although the physics involved is much more complex than for the first resonance, because of electronic correlations. Fig. \ref{cotunneling} shows the transport spectroscopy for the second and third resonances. In addition to the lines of negative differential conductance due to the LDOS fluctuations, a clear line of positive conductance is detected 8\,meV above the ground state. This is lower than the 21\,meV expected for the bulk \cite{Kohn55}, but much larger than in surface donor states measured in ref. \onlinecite{Delft08}.
The line is unambiguously identified as an excited state of the donor by the presence of an inelastic cotunneling line in the Coulomb blockade region \cite{defranceschi}. The third resonance shows features correlated to the second one, proving that it corresponds to the double occupancy of the same donor \cite{defranceschi}: the inelastic cotunneling lines are continued by lines due to transfer $2e^{-} \rightarrow 1e^{-}$ from the lower energy state to the drain. The charging energy is then  $\approx $18\,meV, which, added to the 8\,meV level spacing gives an addition energy for the second electron of 26\,meV. This charging energy is comparable to the 32\,meV deduced in ref. \onlinecite{Sellier06}.
As expected the tunneling rates are very high: the second resonance is well fitted below T=2\,K by a Lorentzian profile
(Breit-Wigner resonance) of intrinsic width $h \Gamma = k_{B}\times$2K, and the conductance at resonance is $\approx$ 0.6 $e^{2}/h$, making cotunneling well visible. The small lines inside the Coulomb diamond are due to elastic cotunneling. This effect is proportionnal to both tunneling rates to source and drain and fluctuates with bias due to LDOS variations.

In conclusion, we have observed a doubling of the ionization energy of a single As donor because of dielectric confinement near an oxide interface. This measurement on a single isolated dopant requires extremely small samples, a condition achieved in ultra-scaled microelectronics devices. This shift in ionization energy is responsible for the very large off-state current up to room temperature, and is therefore a major source of variability for the off-state current in ultra scaled CMOS devices.
Since dopants closer to the gate are much less affected by this shift, gate-all-around structures are the most suitable choice to avoid unacceptable sample to sample variations. On the other hand the next step for taking advantage of this effect is to manipulate single donor electronic orbitals to implement new quantum functionalities in silicon CMOS devices.

\begin{figure}
\begin{center}
\includegraphics[width=0.6\columnwidth]{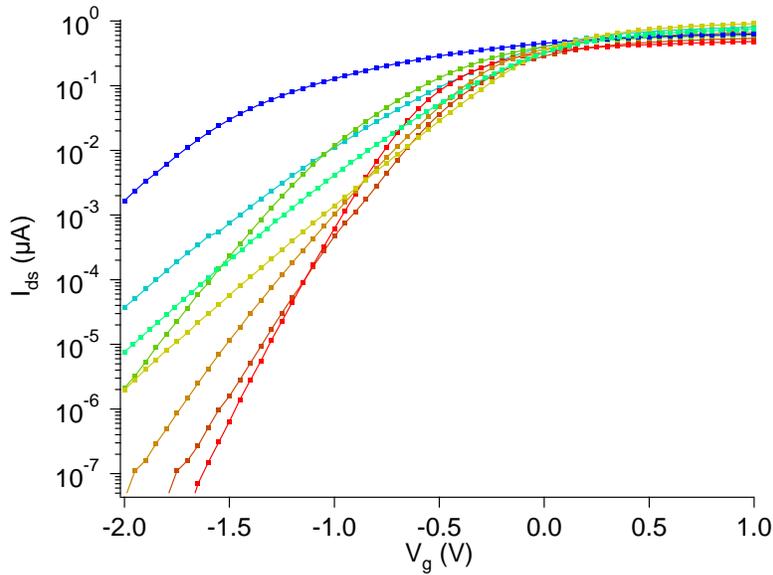}
\caption{figure S1: Room temperature characteristics (drain-source current versus gate voltage) for a set of nominally identical samples from the same wafer. Sample to sample variations are very important, especially below the threshold voltage (-0.5\,V for the best device, in red). In the most extreme case (blue curve) the many decades of excess current compared to the best device are attributed to direct transport through individual arsenic dopants which have diffused into the channel after doping and annealing steps. The source-drain voltage is 10\,mV for these data.}
\label{figS1}
\end{center}
\end{figure}

\begin{figure}
\begin{center}
\includegraphics[width=0.6\columnwidth]{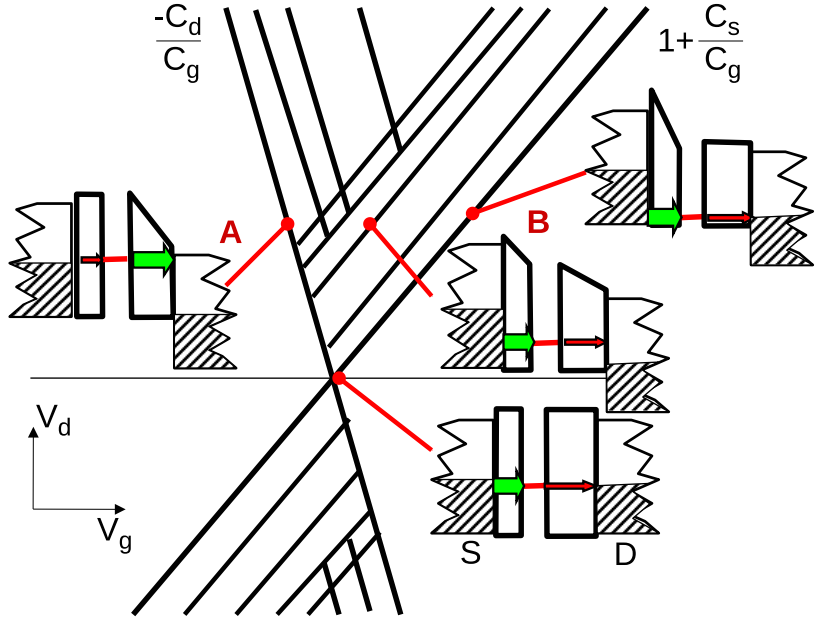}
\caption{figure S2: Sketch for negative or positive differential conductance lines arising from the spectroscopy of the source and drain local density of states (LDOS) probed by a single dopant level. A single dopant level (red line) has a slightly unbalanced tunneling rate to source and drain: the barrier to drain is less transmissive at zero drain bias, represented by a thicker barrier. The drain source differential conductance is dominated by the less transmissive barrier (red arrow, green arrow for the more transmissive ones) and proportional to the LDOS in the corresponding reservoir.  At finite bias around point B the lines are due to LDOS fluctuations in the drain, because the drain barrier is less transmissive. On the opposite, at finite bias in A the drain barrier becomes more transmissive than the source barrier due to the electric field (Fowler-Nordheim tunneling): differential conductance lines are then proportional to the LDOS fluctuations in the source.}
\label{figS2}
\end{center}
\end{figure}

\end{document}